\documentclass[aps,
twocolumn,
  preprintnumbers,amsmath,amssymb]{revtex4}

\usepackage{graphicx}
\usepackage{xcolor}

\begin{document}
\title{Nonergodic Brownian oscillator: High-frequency response}
\author{Alex V. Plyukhin}
\email{aplyukhin@anselm.edu}
 \affiliation{ Department of Mathematics,
Saint Anselm College, Manchester, New Hampshire 03102, USA 
}

\date{\today}

\begin{abstract}
  We consider a Brownian oscillator whose coupling to the environment
  may lead to the formation
  of a localized normal mode. For lower values of the oscillator's  
  natural frequency,
  $\omega\le\omega_c$, the localized mode is absent and
  the unperturbed oscillator reaches thermal equilibrium.
  For higher
  values of $\omega>\omega_c$,  when the localized mode is formed, the unperturbed
  oscillator
  does not thermalize but rather evolves into a nonequilibrium cyclostationary state.
  We consider the response 
  of such an oscillator to an external periodic force. 
  Despite the coupling to the environment, the oscillator
  shows the unbounded resonance (with  the response linearly increasing with time) 
  when the frequency of the external force coincides with
  the frequency of the localized mode.  
  An unusual resonance (``quasi-resonance")
  occurs for the oscillator with the critical value of the natural frequency
  $\omega=\omega_c$, which separates thermalizing (ergodic) and
  non-thermalizing (nonergodic) configurations.
In that case the resonance response increases with time sublinearly, which can be 
interpreted as a resonance between the external force and 
the incipient localized mode.
\end{abstract}

\maketitle

 \section{Introduction}
 Wave localization often occurs,
as in Anderson localization,   
due to destructive interference of waves
from multiple scatterers, but
it also can be caused by a single defect of mass or potential
in extended periodic structures~\cite{Montroll,Takeno,Kashiwamura,Rubin,Todo,Heat}. 
Effects of localized modes on dynamics of 
the classical Brownian (open) oscillator
were addressed, to the best of our knowledge, only relatively 
recently, using the formalism of 
the generalized Langevin equation~\cite{Kemeny,Dhar,Plyukhin}.
Following Ref.~\cite{Plyukhin}, we will refer
to an open oscillator, 
whose coupling to the thermal bath may generate a localized mode, as the nonergodic
Brownian oscillator. In the presence of a localized mode, 
the oscillator does not reach 
thermal equilibrium with the bath but
evolves into a  
cyclostationary state in which  
the mean values and correlations of dynamical variables oscillate with the frequency
of the localized mode. 
Cyclostationary stochastic processes are not  stationary
and, therefore,  manifestly nonergodic.

Compared to other mechanisms of the ergodicity 
breaking~\cite{Morgado,Costa,Bao,Lapas,Deng,Siegle},
the formation of localized modes 
is easier to connect to specific, albeit often idealized, physical models. 
In most of these models the thermal bath is represented by a lattice~\cite{Montroll,Takeno,Kashiwamura,Rubin,Todo,Heat},
but that does not appear to be necessary. 
It was suggested that wave localization might be important for the functional dynamics of proteins~\cite{Chalopin}.
The presence or absence of a localized mode can often  
be controlled by an experimentally tunable parameter, e.g. the oscillator's natural frequency $\omega$.
For the model discussed in this paper, 
a localized mode is formed when $\omega$ exceeds a certain critical value,
$\omega>\omega_c$.
By varying the oscillator frequency, one can engineer a broader 
class of nonequilibrium processes
 which may involve both ergodic ($\omega\le\omega_c$) 
and nonergodic ($\omega>\omega_c$)
configurations.

The previous studies of the nonergodic oscillator were focused
on its relaxation
and correlation properties in the absence of external forces.
In this paper, we consider
the dynamical response of  a nonergodic 
Brownian oscillator to  
the external harmonic force $F_{ex}(t)=F_0\sin(\Omega\, t)$.
The response has the form of unbounded resonance  when the external frequency $\Omega$ 
equals the frequency of the localized mode. Most interesting  is the response 
of the oscillator with the critical  natural frequency $\omega_c$ just below the
formation of the localized mode. In that case a resonance response will be shown to increase 
with time sublinearly.

\section{Model}
We consider a Brownian oscillator described by the generalized
Langevin equation~\cite{Zwanzig,Weiss} 
\begin{eqnarray}
  \!\!
  \ddot{x}=\!-\!\omega^2 x
  \!-\!\!\int_0^t
\!\!
  K(t\!-\!\tau)\,\dot{x}(\tau) d\tau
  \!+\!\frac{F_0}{m}\sin(\Omega t)
  \!+\!\frac{1}{m}\,\xi(t),
\label{GLE}
\end{eqnarray}
where the noise $\xi(t)$ is zero-centered and connected 
to the dissipation kernel 
$K(t)$
by the standard fluctuation-dissipation relation.
The generalized Langevin equation can be rigorously derived from
first principles
and, in contrast to its Markovian (time-local) counterpart,  may hold
on the time scale
comparable  with the relaxation time of the thermal bath.
The latter is important for systems (particularly, viscoelastic) 
with a broad hierarchy of relevant time scales~\cite{Xie1,Xie2,Goychuk}.

We consider a specific dissipation kernel
\begin{eqnarray}
  K(t)=\frac{\omega_0^2}{4}\,[J_0(\omega_0t)+J_2(\omega_0t)]
    =\frac{\omega_0}{2}\,\frac{J_1(\omega_0 t)}{t},
\label{K}
\end{eqnarray}
where $J_n(x)$'s are Bessel functions of the first kind.
The kernel has the absolute maximum at $t=0$ and for $t>0$
it oscillates with an amplitude decaying with time as $t^{-3/2}$.
For $F_0=0$ and $\omega=0$, the generalized Langevin equation with kernel (\ref{K}) 
describes
Brownian motion of the terminal atom of a semi-infinite harmonic chain,
which is a version of Rubin's model~\cite{Weiss}.

The special feature of the kernel (\ref{K}) is that  its 
spectral density $\rho(\nu)$ has a finite upper bound $\omega_0$,
  \begin{eqnarray}
    \rho(\nu)\!=\!\int_0^\infty\!\!\! K(t)\cos(\nu t) \,dt
    \!=\!\frac{1}{2}\sqrt{\omega_0^2-\nu^2}
  \,\,\theta(\omega_0-\nu),
\label{rho}
\end{eqnarray}
where $\theta(x)$ is the step function.
That is known to be  a condition for
the formation of a localized mode whose frequency $\omega_*$ lies 
outside the spectrum, $\omega_*>\omega_0$~\cite{Montroll,Takeno,Kashiwamura,Rubin}.
Thus, the unperturbed oscillator has three characteristic
frequencies, $\omega$, $\omega_0$, and $\omega_*$.
The first two, $\omega$ and $\omega_0$, are explicitly present in the Langevin equation,
whereas the third $\omega_*$ is not, and can be viewed 
as a hidden parameter. One may expect a singular response  when
the external frequency $\Omega$ coincides with (or is close to) one of the three
characteristic frequencies. 
Since the localized mode frequency lies in the interval
$\omega_*>\omega_0$, the model shows most interesting results 
for the high frequency response at 
$\Omega\ge\omega_0$.  We will limit ourselves to that case.
The response properties at lower frequencies $\Omega<\omega_0$ require a 
somewhat different mathematical approach 
(see a remark in the Conclusion) 
and will be considered elsewhere.

As far as only the first moments  of the coordinate
and its derivatives
are concerned, the
stochastic nature of the generalized Langevin equation
and the fluctuation-dissipation
relation are redundant. By averaging Eq. (\ref{GLE}) one 
gets for the average
displacement $q(t)=\langle x(t)\rangle$
the integro-differential equation
\begin{eqnarray}
  \!\!
  \ddot{q}=-\omega^2 q-\int_0^tK(t-\tau)\,\dot{q}(\tau)\,d\tau+\frac{F_0}{m}
  \,\sin(\Omega t),
\label{q_eq}
\end{eqnarray}
which is totally deterministic.
This equation was the subject of several recent studies, particularly
for the case of the fractional oscillator with a power-law
dissipation kernel
$K(t)\sim t^{-\alpha}$ \cite{Barkai1,Barkai2}. 
Solutions, while showing a number of new interesting features, 
were still found 
to satisfy general expectations of the linear response theory and 
typical experimental setups: 
They involve transient terms which die out at long times and 
a 
steady-state solution which oscillates with the frequency of the
external field and a time-independent amplitude. As shown below, for 
a nonergodic oscillator the solution may have a very different structure.

We  assume that for  $t<0$ the external force is zero
and the oscillator
at $t=0$
is in thermal equilibrium with the bath. This implies
zero initial conditions $q(0)=\dot q(0)=0$.
Then the solution of Eq. (\ref{q_eq})
in the Laplace domain reads
\begin{eqnarray}
  \tilde q(s)=\frac{1}{m}\,\tilde G(s)\,\tilde F_{ex}(s),
  \label{sol_laplace}
  \end{eqnarray}
where the Laplace transforms of the Green's function $G(t)$ and the external force are
\begin{eqnarray}
  \tilde G(s)=\frac{1}{s^2+s\,\tilde K(s) +\omega^2}, \quad
     \tilde F_{ex}(s)=F_0\,\frac{\Omega}{s^2+\Omega^2}.
  \label{aux1}
  \end{eqnarray}
 In the time domain, solution Eq. (\ref{sol_laplace}) has the form of the convolution,
\begin{eqnarray}
  q(t)=\frac{F_0}{m}\,\int_0^t G(t-\tau)\,\sin(\Omega\,\tau)\,d\tau.
\label{response_conv}
  \end{eqnarray} 
This expression is general; peculiarities of the model reside in 
 the specific form of the Green's function $G(t)$.
Substituting the Laplace transform of kernel (\ref{K}),
\begin{eqnarray}
  \tilde K(s)=\frac{1}{2}\, \left(
  \sqrt{s^2+\omega_0^2}-s\right)
\end{eqnarray}
into Eq. (\ref{aux1}) for $\tilde G(s)$ one gets
\begin{eqnarray}
\tilde G(s)=\frac{2}{s^2+s\,\sqrt{s^2+\omega_0^2}+2\,\omega^2}.
\label{G2}
\end{eqnarray}
The inversion of this  transform can be expressed in terms of standard functions only for
special values of the oscillator frequency $\omega$, see Eq. (\ref{special}) below. 
For arbitrary $\omega$,
the Green's function in the time domain $G(t)$ can be expressed in an integral form 
inverting  $\tilde G(s)$ by evaluating
a relevant Bromwich integral in the complex plane.
As shown in Ref.~\cite{Plyukhin}, for
the given model
there is  a critical value of the oscillator frequency 
\begin{eqnarray}
  \omega_c=\omega_0/\sqrt{2}\approx 0.707\,\omega_0,
\end{eqnarray}
which separates two types of the system's  behavior,
\begin{eqnarray}
G(t)=
\begin{cases}
    G_e(t), & \text{if $\omega\le\omega_c$},\\
G_e(t)+G_0\,\sin(\omega_* t), & \text{ if $\omega>\omega_c$}.
  \end{cases}
\label{G_gen}
\end{eqnarray}
For $\omega\le\omega_c$ a localized mode is not formed, 
and the Green's function involves only the ergodic component
 \begin{eqnarray}
   G_e(t)=\frac{4}{\pi\omega_0}
   \int_0^1 \frac{\sin(x\,\omega_0\,t)\,x\,\sqrt{1-x^2}\,dx}
{(1-4\,\lambda^2)\,x^2+4\,\lambda^4},
\label{Ge}
\end{eqnarray}
where $\lambda$ denotes the dimensionless oscillator frequency in units of $\omega_0$.
We will also use the notation $\lambda_c$ for the dimensionless critical oscillator frequency,
\begin{eqnarray}
\lambda=\omega/\omega_0,\qquad \lambda_c=\omega_c/\omega_0=1/\sqrt{2}.
\label{lambda}
\end{eqnarray}
One can verify that $G_e(t)$ for any $\lambda$ 
non-monotonically and slowly decreases and vanishes at long times.
We will refer to settings with $\omega\le\omega_c$ ($\lambda\le\lambda_c$) as ergodic 
configurations. 
One can show that the oscillator in ergodic configurations reaches 
thermal equilibrium at long times~\cite{Plyukhin}.
The Green's function $G(t)$ has also the meaning of the 
(normalized) correlation function $\langle x(0) \dot x(t)\rangle$ ~\cite{Plyukhin}.
The asymptotic behavior $G_e(t)\to 0$  
corresponds to the asymptotic fading of correlations and relaxation 
to thermal equilibrium.

For  $\omega>\omega_c$, as Eq. (\ref{G_gen}) shows, the localized mode is developed,
and the Green's function involves
both ergodic and nonergodic components. 
The latter does not vanish at long times but rather 
oscillates with the localized mode frequency $\omega_*$.
The localized mode amplitude $G_0$ and frequency $\omega_*$
of  the nonergodic component
are given by the following expressions~\cite{Plyukhin}:
\begin{eqnarray}
  \omega_0\,G_0=\frac{8\,\lambda^2-4}{(4\,\lambda^2-1)^{3/2}}, \qquad
 \omega_*=\frac{2\,\lambda^2}{\sqrt{4\lambda^2-1}}\,\,\omega_0.
  \label{omega_loc}
\end{eqnarray}
In settings with $\omega>\omega_c$ ($\lambda>\lambda_c$), which we refer to as nonergodic configurations, correlations do not vanish at long times.
The oscillator does not reach thermal equilibrium,
but evolves to a cyclostationary state whose statistics oscillate with frequency $\omega_*$.

For two values of the oscillator frequency, $\omega=\omega_0/2$ and $\omega=\omega_c$, 
both corresponding to ergodic configurations,
the inverse transform
of Eq. (\ref{G2}), or the integral expression (\ref{Ge}), can be compactly expressed in terms of Bessel functions,
\begin{eqnarray}
G(t)=G_e(t)=
\begin{cases}
    \frac{8}{\omega_0^2 t}\,J_2(\omega_0 t) &\text{if}\quad\omega=\omega_0/2,\\
\frac{2}{\omega_0}\,J_1(\omega_0 t) &\text{if}\quad\omega=\omega_c.
  \end{cases}
  \label{special}
\end{eqnarray}
One observes that for the critical configuration ($\omega=\omega_c$) the Green's function
decays slower. That feature can be viewed as a precursor of the localized mode formation and
leads to conspicuous response properties.

\section{Response of critical configuration: Quasi-resonance}
The most appealing type of response,
which we refer to as  quasi-resonance, 
occurs when 
the oscillator frequency has the critical value and 
the external force frequency is equal to the cutoff frequency of the bath 
spectrum,  
\begin{eqnarray} 
  \omega=\omega_c, \qquad 
  \Omega=\omega_0.
  \label{case2_condition}
\end{eqnarray}
The Green's function, according to Eq. (\ref{special}), takes the form
$G(t)=\frac{2}{\omega_0}\,J_1(\omega t)$. Substituting it 
into Eq. (\ref{response_conv}) and taking into account 
Kapteyn's integral
\begin{eqnarray} 
  \int_0^t J_1(\tau)\,\sin(t-\tau)\,d\tau=\sin(t) -t J_0(t),
  \label{Kapteyn}
\end{eqnarray}
see Ref.~\cite{Watson} and the Appendix,
immediately yields
\begin{eqnarray} 
  q(t)=\frac{2\,F_0}{m\,\omega_0^2}\,\Big\{\sin(\omega_0 t)
  -\omega_0t\,J_0(\omega_0t)
  \Big\}. 
\label{case2}
\end{eqnarray}
Here the first term is the anticipated steady-state solution
oscillating with the frequency of the driving force $\Omega=\omega_0$
(remarkably, with  a zero phase shift).
The second term, however, is quite unexpected. 
Instead of being transient, it oscillates with an amplitude
increasing indefinitely in time as $\sqrt{t}$.
Such a resonance-like (quasi-resonance) behavior is in drastic contrast to that of the normal
damped and fractional oscillators when 
the resonance solution is stationary, and its 
amplitude is finite.

One may view the configuration with $\omega=\omega_c$ as a critical phase
where the localized mode is incipient and its frequency
coincides with the cutoff frequency of the bath spectrum,
$\lim_{\lambda\to\lambda_c}\omega_*=\omega_0$, see Eq. (\ref{omega_loc}).
Unperturbed properties of such a phase
show no signs of any anomalies, except a slower decay of the Green's function.
However, the dynamical response to the force with
the frequency of the incipient 
localized mode, $\Omega=\omega_0$, is singular.
One might suggest the following interpretation. Recall that the Green's function is also
the correlation function $\langle x(0)\dot x(t)\rangle$. The slower decay of 
correlations in the critical configuration signifies 
the  slower heat exchange 
between the system and the heat bath. As a result, the system receives energy
from the external source with the rate higher than
the rate of heat dissipation into the heat bath, which makes 
the response to increase with time indefinitely.
This interpretation, however,
is somewhat superficial and 
does not fully catch the subtlety of the result. It 
does not explain the sublinear increase of the response with time.
Also, applying the similar reasoning to the normal damped oscillator,
one might expect that for a sufficiently small dissipation coefficient
the resonance response would increase with time indefinitely.
That, however, is not the case.

\section{Response of special ergodic configuration}
Another special setting when the response can be expressed in a compact analytical 
form is $(\omega=\omega_0/2,\,\Omega=\omega_0)$.
The Green's function is given by the first expression
in Eq. (\ref{special}), which 
can also be presented as
\begin{eqnarray} 
G(t)=G_e(t)=\frac{2}{\omega_0}\left[
J_1(\omega_0 t)+J_2(\omega_0 t)\right].
\label{aux2}
\end{eqnarray}
Substituting this into Eq. (\ref{response_conv}) and taking
into account integral (\ref{Kapteyn}) 
and its generalization~\cite{Bailey}  
\begin{eqnarray} 
\!\!\!
\int_0^t \!\! J_3(\tau)\,\sin(t\!-\!\tau)\,d\tau
=-t \, J_2(t)\!+
\!6J_1(t)\!-\!3\sin(t),
\end{eqnarray}
which also can be expressed as 
\begin{eqnarray} 
\!\!\!
\int_0^t\!\! J_3(\tau)\,\sin(t\!-\!\tau)\,d\tau
\!=\!t J_0(t)\!+\!4J_1(t)\!-\!3\sin(t),
\end{eqnarray}
yields
\begin{eqnarray} 
  q(t)
  =\frac{4\,F_0}{m\,\omega_0^2}\,\Big\{
  \sin(\omega_0 t-\pi)+
  2J_1(\omega_0t)\Big\}.
\label{case1}
\end{eqnarray}
The structure of this solution is similar to that
for the normal damped oscillator.
The first term is the steady-state solution which oscillates with the frequency of the external
force $\Omega=\omega_0$.
The second  term is a transient vanishing at long times.
 Notable features are as follows: (1) a slow decay of the transient term,
and (2) the phase shift $\pi$ of the steady-state term is the same as for the undamped
oscillator. Recall that for the normal damped oscillator
the phase shift reaches the value $\pi$ only in the limit $\Omega\to\infty$.

\section{Response of general 
ergodic configurations} 
Let us consider the response 
of general ergodic (subcritical) configurations
with $\omega\le\omega_c$. 
The Green's function has only  an ergodic component, $G(t)=G_e(t)$.
Substituting Eq. (\ref{Ge}) for $G_e(t)$ into Eq. (\ref{response_conv}),
changing the integration order, and integrating over $\tau$ 
one obtains
\begin{eqnarray}
  q(t)=\frac{F_0}{m\omega_0^2}\Big\{
  -A\,\sin(\Omega t)+\varphi(t)\Big\},
\label{qe}
\end{eqnarray}
where the amplitude $A$ and transient $\varphi(t)$, both dimensionless,  
are given by the integral expressions
\begin{eqnarray}
  \!\!\!\!\!
  A&=&\frac{4}{\pi}
  \int_0^1
  \!\!
  \frac{x^2\sqrt{1-x^2}\,dx}
      {
    \Big[(1-4\lambda^2)\,x^2+4\lambda^4\Big]
   \left(\Lambda^2-x^2\right) 
      },
      \label{amp}\\
 \!\!\!\!\!\varphi(t)&=&\frac{4\Lambda}{\pi}\!
 \int_0^1
 \!\!
 \frac{x\sqrt{1-x^2}\,\sin(x\,\omega_0\,t)\,dx}
      {
    \Big[(1-4\lambda^2)\,x^2+4\lambda^4\Big]
   \left(\Lambda^2-x^2\right) 
      }.
      \label{varphi}
\end{eqnarray}
Here $\lambda$ stands, as above, for the dimensionless oscillator frequency, 
and $\Lambda$ denotes the dimensionless external force frequency, both in
units of the cut-off frequency $\omega_0$,
\begin{eqnarray}
  \lambda=\omega/\omega_0, \quad \Lambda=\Omega/\omega_0.
\end{eqnarray}
The considered domain $(\omega\le\omega_c, \,\Omega\ge\omega_0)$ corresponds to
$(\lambda\le\lambda_c, \,\Lambda\ge 1)$.

For the strict inequality $\Lambda>1$, 
the integrands in the above expressions have no singularities, so the integrals converge. 
For $\Lambda=1$ and $\lambda\ne \lambda_c$, the integrands have an integrable singularity
at the upper integration limit, and the integrals still converge.
For the special case $(\lambda=\lambda_c,\,\Lambda=1)$ the integrals 
diverge at the upper integration limit,
and the above expressions are not valid.
That case, however, was already described in Sec. III by another method.

Excluding the special case $(\lambda=\lambda_c,\,\Lambda=1)$,
the integral expression    
Eq. (\ref{amp}) for the amplitude $A$ can  be worked out to
the explicit form
\begin{eqnarray}
  A=\frac{2}{\Lambda^2+\Lambda\sqrt{
      \Lambda^2-1}-2\lambda^2}.
      \label{amp2}
\end{eqnarray}
Equation (\ref{varphi}) for $\varphi(t)$
is reduced to a more explicit form
apparently only for $(\lambda=1/2, \, \Lambda=1)$,  
which is one of the two special settings considered above. 
For that case, Eqs. (\ref{varphi}) and (\ref{amp2})
give $\varphi(t)=8J_1(\omega_0 t)$ and $A=4$,
and Eq. (\ref{qe})  recovers Eq. (\ref{case1}).

According to Eq. (\ref{amp2}),
for the considered domain $(\lambda\le\lambda_c, \Lambda\ge 1)$
the amplitude $A$ is  
positive. Then we can write the result (\ref{qe}) as
\begin{eqnarray}
  q(t)=
  \frac{F_0}{m\omega_0^2}\Big\{
  A\,\sin(\Omega t-\pi)+\varphi(t)\Big\}.                        
  \label{qe2}
\end{eqnarray}
Recall again that  the phased shift $\pi$  
is a property similar to that of the {\it undamped} oscillator
for $\Omega>\omega$, whereas  for the normal damped oscillator
the phase shift reaches  $\pi$ only in the limit $\Omega\gg \omega$.

\begin{figure}[t]
  \includegraphics[
    width=9cm,
    height=9cm]{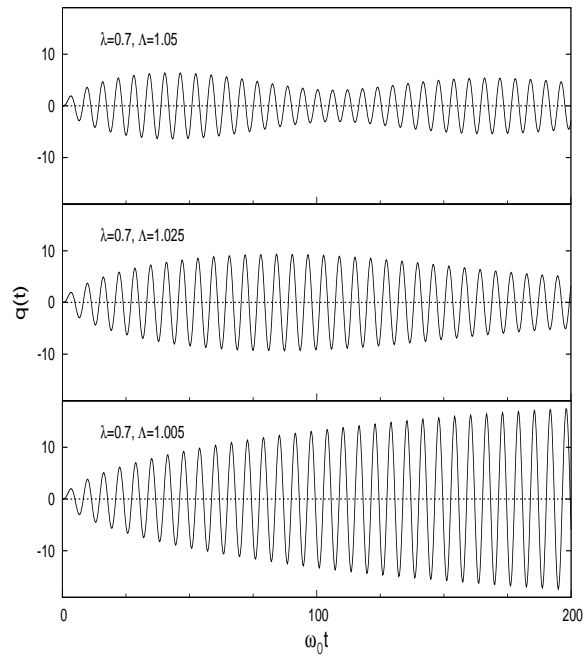}
\caption{The average coordinate $q(t)$, 
in units of $q_0=F_0/(m\omega_0^2)$,
  for ergodic configurations with the dimensionless oscillator
  frequency $\lambda=\omega/\omega_0=0.7$ (just below the critical
  value $\lambda_c=1/\sqrt{2}\approx 0.707$) 
and three values of the dimensionless external frequency 
$\Lambda=\Omega/\omega_0$. At longer times the beat pattern dies out and
$q(t)$ takes the steady-state form Eq. (\ref{sol5}).  
The plot at the bottom is close to that
for the quasi-resonance solution Eq. (\ref{case2}). }
\end{figure}

One can verify that for $\lambda<\lambda_c$ the term $\varphi(t)$
given by Eq. (\ref{varphi})  
is transient and
vanishes at long times. Then the asymptotic solution
is given by the steady-state term, oscillating with the frequency
of the external force,
\begin{eqnarray}
  q(t)\to\frac{F_0\,A}{m\omega_0^2}\,\sin(\Omega t-\pi) \quad
  \text{as}\quad t\to\infty.
\label{sol5}
\end{eqnarray}
However, the decays of the transient
$\varphi(t)$ is slow,
and it is getting slower as $\lambda\to\lambda_c$. 
As a result, 
there is a significant time interval,
whose duration increases, in fact diverges,
as $\lambda\to\lambda_c$, when $\varphi(t)$
oscillates with an almost constant amplitude.
Then the solution
$q(t)$ is governed
by the interplay of two oscillating terms in Eq. (\ref{qe2}).
As a result, 
the solution,  during a long, albeit finite, time interval,
has not a harmonic form (\ref{sol5}),
but instead shows a beat pattern.
For a fixed value of $\lambda$ close  to $\lambda_c$, 
the beat period tends to increase when $\Lambda\to 1$.
In the limits $\lambda\to\lambda_c$ and $\Lambda\to 1$
the initial increasing section
of the first beat has the infinite duration, 
and the solution takes the quasi-resonance form
Eq. (\ref{case2}). The tendency is shown in Fig. 1.

\section{Response of nonergodic configurations} 
Consider now the
oscillator with natural frequency $\omega>\omega_c$.
The localized mode now is
  fully developed, and the Green's function, according to  
  Eq. (\ref{G_gen}), involves  the harmonic (nonergodic)
  term $G_0\sin\omega_*t$. In that case
  one may anticipate the response to be similar
  to that of the undamped oscillator with the natural frequency $\omega_*$,
  showing the resonance at $\Omega=\omega_*$.
  The expectation is confirmed by the calculations below.  

For nonergodic configurations with $\omega>\omega_c$
the Green's function has now both ergodic
and nonergodic (periodic) components,
$G(t)=G_e(t)+G_0\,\sin(\omega_* t)$.
Substituting this into Eq. (\ref{response_conv}),
taking into account Eq. (\ref{Ge})
for $G_e(t)$, 
changing the integration order, and integrating over $\tau$ 
yields the result for  $q(t)$. We write it as
\begin{eqnarray}
  q(t)=q_e(t)+q_{ne}(t),
\label{q5}
  \end{eqnarray}
where the first and second terms come from the ergodic
and nonergodic components
of the Green's function, respectively.
The ergodic term $q_e(t)$ coincides with the response
of the ergodic configuration
given by Eq. (\ref{qe2}), 
\begin{eqnarray}
  q_e(t)=
  \frac{F_0}{m\omega_0^2}\Big\{
  A\,\sin(\Omega t-\pi)+\varphi(t)\Big\},                       
  \label{qe3}
\end{eqnarray}
where $A$ and $\varphi(t)$ are still given by integral expressions
(\ref{amp}) and (\ref{varphi}).
However, for the given domain $(\lambda>\lambda_c,\,\Lambda\ge1)$,
expression (\ref{amp}) for $A$ 
is reduced not to Eq. (\ref{amp2}),
but to a more involved form
\begin{eqnarray}
  A=\frac{2 (4\lambda^2-1)\left(\Lambda^2-\Lambda\sqrt{\Lambda^2-1}\right)-4\lambda^2}
  {(4\lambda^2-1)(4\lambda^4-4\lambda^2\Lambda^2+\Lambda^2)}.
  \label{amp3}
\end{eqnarray}
At $\Lambda=\Lambda_0=4\lambda^4/(4\lambda^2-1)$
both the numerator and the denominator are zero,
and the expression has to be extended by continuity, 
\begin{eqnarray}
  A(\Lambda_0)=\lim_{\Lambda\to\Lambda_0} A=\frac{1}{2\lambda^2\,
  (4\lambda^2-1)\,(2\lambda^2-1)}.
  \label{amp4}
\end{eqnarray}
One observes that the amplitude $A$
behaves qualitatively similar to that for ergodic configurations,
Eq. (\ref{amp2}).
For any fixed value of $\lambda>\lambda_c$
 the amplitude $A$ as a function of $\Lambda$ monotonically decreases
 and shows no maximum (no resonance)
 near $\Lambda=\lambda$ ($\Omega=\omega$). 
 
 The function $\varphi(t)$ in Eq. (\ref{qe3}) is still
 given by Eq. (\ref{varphi}).
One can verify numerically that for the given domain it is
transient, i.e. dies out at long times. However,
as for the subcritical case
$\lambda\le\lambda_c$, the decay time of $\varphi(t)$
is getting longer and diverges when $\lambda\to\lambda_c$ and $\Lambda\to 1$.
As a result, the ergodic component $q_e(t)$ behaves
similar to the solution for ergodic configurations with $\lambda\le\lambda_c$.
For $\lambda$ close to $\lambda_c$
and $\Lambda$ close to $1$,
$q_e(t)$ shows on a shorter time scale beats patterns similar to those
illustrated in Fig. (1).
In the limit
$\lambda\to\lambda_c^+$ and $\Lambda\to 1^+$ the duration of the first beat
diverges and  $q_e(t)$ takes the
quasi-resonance form (\ref{case2}).
For a finite $\lambda-\lambda_c>0$, $q_e(t)$ evolves on a longer time scale  
to the steady
state solution oscillating with frequency $\Omega$.

\begin{figure}[t]
\includegraphics[width=9cm,
    height=9cm]{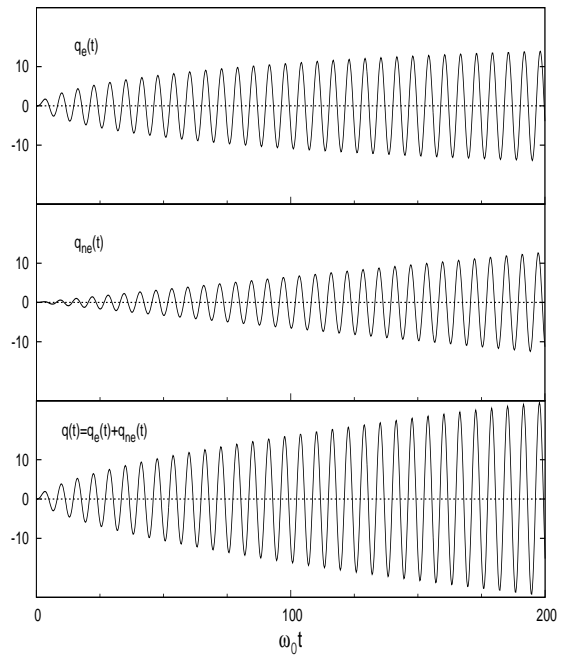}
\caption{The response functions for a nonergodic configuration with $\lambda=0.72$ and 
$\Lambda=1.005$. Top: the ergodic component $q_e(t)$, Eq. (\ref{qe3}). Middle: the nonergodic component 
  $q_{ne}(t)$, Eq. (\ref{qne1}). Bottom: the total solution, $q(t)=q_e(t)+q_{ne}(t)$.
The ergodic and nonergodic components oscillate with amplitudes which
increase with time as $\sqrt{t}$ and $t$,
respectively.  
}
\end{figure}

Consider now the the nonergodic term $q_{ne}(t)$ in Eq. (\ref{q5}).
Substituting the nonergodic component of the Green's function
$G_0\sin\omega_* t$
into Eq. (\ref{response_conv}) yields
\begin{eqnarray}
  \!\!\!
  q_{ne}(t)\!=\!\frac{F_0\,G_0}{m}\,\frac{1}{\Omega^2-\omega_*^2}\Big[
    \Omega\,\sin(\omega_* t)\!-\!\omega_*\sin (\Omega \,t)
    \Big],
  \label{qne1}
\end{eqnarray}
for $\Omega\ne\omega_*$, and
\begin{eqnarray}
  q_{ne}(t)=\frac{F_0\,G_0}{2m}\,\Big[
    \frac{1}{\Omega}\,\sin(\Omega\, t)-t\,\cos(\Omega\, t)
    \Big],
  \label{qne2}
\end{eqnarray}
for $\Omega=\omega_*$. Those are exactly  the expressions for the
response of the undamped oscillator with the natural
frequency $\omega_*$,
describing  beat patterns for  $\Omega\ne\omega_*$
and  unbounded resonance for $\Omega=\omega_*$.


The total response $q(t)=q_e(t)+q_{ne}(t)$ 
is determined by the interplay of both ergodic and nonergodic components
and shows a variety of beat patterns for $\Omega\ne \omega_*$ and 
unbounded resonance for $\Omega=\omega_*$.
Of special interest is the asymptotic
case ($\lambda\to\lambda_c^+,\, \Lambda\to 1^+$)
when both components  on a long time scale show the resonance-like behavior,
$q_e(t)\sim t J_0(\omega t)$ and $q_{ne}(t)\sim t\cos(\omega_0 t)$.
The case is illustrated in Fig. 2.

\section{Conclusion}
The response properties of an open oscillator
with a well-developed localized mode with frequency $\omega_*$
are similar to those of an {\it isolated}
oscillator with the natural frequency $\omega_*$.
In particular, when
the frequency of the external force $\Omega$ coincides with $\omega_*$,
the oscillator, instead of evolving into a steady state, 
shows an unbounded resonance.
Superficially, this might come as a surprise since
the equation of motion (\ref{q_eq}) involves a dissipation term
(which usually smooths out resonance singularities) 
and  also because the equation does not involve the
frequency $\omega_*$ explicitly.
However, from a more educated point of view, 
which we tried to develop in this paper, 
the unbounded resonance at $\Omega=\omega_*$ is hardly
unexpected, considering that the localized mode does not exchange energy
with the thermal bath and thus behaves as an isolated oscillator.

\begin{figure}[t]
  \includegraphics[
    width=8cm,
    height=8cm]{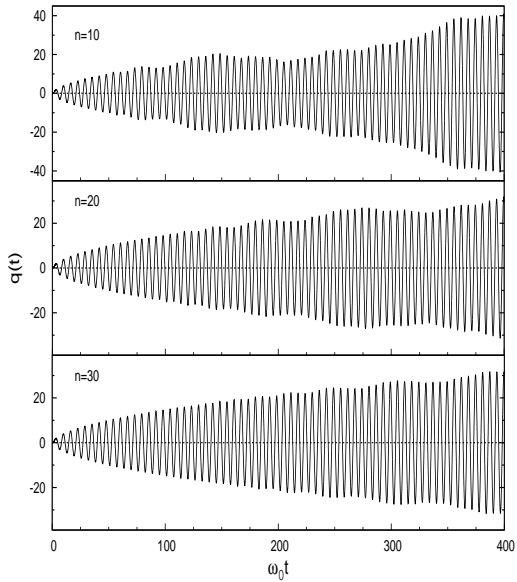}
\caption{Simulation results 
for the quasi-resonance set up
($\omega=\omega_c$, $\Omega=\omega_0$) when the bath is represented by a finite 
harmonic chain of $n=10, 20$, and $30$  atoms. As $n$ increases further, 
the simulation results quickly converges to Eq. (\ref{case2}).}
\end{figure}

More subtle is the result for the critical value
of the oscillator natural frequency $\omega_c$
when, according to Eq. (\ref{G_gen}), the localized mode is incipient.
In that case, even though the localized mode
does not affect characteristics of the unperturbed
system, the dynamical response may have
a singular quasi-resonance form (\ref{case2}), which 
has no analog or counterpart in other 
open oscillator models. 
While only a specific dissipation kernel (\ref{K}) was considered here,
one may expect similar results for other kernels whose 
spectral density has a finite upper bound.

Although the presented results are exact, it might be of interest to verify and extend them with numerical simulations. As we already mentioned, the generalized Langevin equation with the kernel (\ref{K}) describes
a terminal atom of a semi-infinite harmonic chain
subjected to an external harmonic potential and driven by an external periodic force. 
Figure 3 shows the simulation results for the quasi-resonance response of the 
oscillator coupled to the finite chain of $n$ atoms. The dependence of the response on 
the size of the thermal bath may be of interest for biochemical applications
when both a system and a bath correspond to degrees of freedom of a single macromolecule~\cite{Chalopin,Xie1,Xie2}.
Simulation shows that for $n$ of an order of $100$ or more 
the quasi-resonance response is practically indistinguishable from the result (\ref{case2}) for the infinite bath. For smaller $n$ the amplitude of oscillations as a function of time is non-monotonic, 
yet on the long run the response increases with time indefinitely.

We have already noted in Ref.~\cite{Plyukhin} that
the parametric transition between ergodic ($\omega\le \omega_c$)
and nonergodic ($\omega>\omega_c$) configurations resembles
a phase transition of the second kind.
From that perspective, the quasi-resonance response
of the configuration with $\omega=\omega_c$
can be viewed as a critical phenomenon, and 
the exponent $1/2$ in the asymptotic form
of Eq. (\ref{case2}), 
$q(t)\sim t^{1/2}\cos(w_0t-\pi/4)$,  
can be interpreted as a critical exponent.

In this paper we considered the response  only at high frequency $\Omega\ge\omega_0$. For the low-frequency response at $\Omega<\omega_0$, or $\Lambda< 1$, the integral expressions (\ref{amp}) and (\ref{varphi}) for $A$ and $\varphi(t)$ diverge and are not valid. One can show that 
the results can be extended for the low-frequency response merely by 
defining the improper integrals in Eqs. (\ref{amp}) and (\ref{varphi}) 
in the sense of Cauchy principal value.
The justification, however, requires a more involved technique and will be addressed  in a future publication.

\renewcommand{\theequation}{A\arabic{equation}}
  \setcounter{equation}{0}  

\section*{APPENDIX}
  The integral (\ref{Kapteyn}) 
\begin{eqnarray} 
  I(t)=\int_0^t J_1(\tau)\,\sin(t-\tau)\,d\tau
\label{A1}
\end{eqnarray}
can be evaluated as follows. The  
Laplace transform of $J_1(t)$ is
\begin{eqnarray} 
 \tilde J_1(s)=\mathcal{L} \{J_1(t)\}=\frac{1}{s^2+s\sqrt{s^2+1}+1},
\end{eqnarray}
and the convolution (\ref{A1}) in the Laplace domain has the form
\begin{eqnarray} 
  \tilde I(s)=\mathcal{L} \{I(t)\}=\frac{1}{(s^2+1)(s^2+s\sqrt{s^2+1}+1)}.
\label{A2}
\end{eqnarray}
In terms of partial fractions it can be written  as 
\begin{eqnarray}
\!\!\!\!\!\!
  \tilde I(s)\!=\!
  \frac{1}{s^2+1}\!-\!\frac{s}{(s^2+1)^{3/2}}
  \!=\!
  \frac{1}{s^2+1}
  \!+\!\frac{d}{ds}\frac{1}{\sqrt{s^2+1}}.
\label{A3}
\end{eqnarray}
Then, using the property
\begin{eqnarray} 
\mathcal{L} \{t\,f(t)\}=-\frac{d}{ds} \mathcal{L}\{f(t)\}=-\frac{d}{ds}\,\tilde f(s)
\end{eqnarray}
and taking account
the transform of the Bessel function
$\mathcal{L}\{J_0(t))=1/\sqrt{s^2+1}$, one finds
\begin{eqnarray} 
  I(t)=\sin(t)-t J_0(t).
\end{eqnarray}
This is Eq. (\ref{Kapteyn}) of the main text.

\end{document}